# Evolution of Griffiths Phase and Critical Behaviour of $La_{1-x}Pb_xMnO_{3\pm y}$ Solid Solutions

Sagar Ghorai[1], Sergey A. Ivanov[1,2], Ridha Skini[1], Peter Svedlindh[1]

[1] Department of Materials Science and Engineering, Uppsala University, Box 534, SE-751 21, Uppsala, Sweden

[2] Semenov Institute of Chemical Physics, Kosygina Street, 4, Moscow, Russia, 119991

## ABSTRACT

Polycrystalline $La_{1-x}Pb_xMnO_{3\pm y}$ ($x = 0.3, 0.35, 0.4$) solid solutions were prepared by solid state reaction method and their magnetic properties have been investigated. Rietveld refinement of X-ray powder diffraction patterns showed that all samples are single phase and crystallized with the rhombohedral structure in the R-3c space group. A second order paramagnetic to ferromagnetic phase transition was observed for all materials. The Griffiths phase (GP), identified from the temperature dependence of the inverse susceptibility, was suppressed by increasing magnetic field and showed a significant dependence on A-site chemical substitution. The critical behaviour of the compounds was investigated near to their Curie temperatures, using intrinsic magnetic field data. The critical exponents ($\beta$, $\gamma$ and $\delta$) are close to the mean-field approximation values for all three compounds. The observed mean-field like behaviour is a consequence of the GP and the formation of ferromagnetic clusters. Long-range ferromagnetic order is established as the result of long-range interactions between ferromagnetic clusters. The magnetocaloric effect was studied in terms of the isothermal entropy change. Our study shows that the material with the lowest chemical substitution ($x = 0.3$) has the highest potential (among the three compounds) as magnetic refrigerant, owing to its higher relative cooling power (258 J/kg at 5 T field) and a magnetic phase transition near room temperature.

# 1. INTRODUCTION

The perovskite manganite oxides have been studied extensively for their colossal magnetoresistance,[1] Jahn–Teller (J-T) distortions,[2] metal-insulator transitions,[3] and most importantly the strong coupling between lattice, charge, orbit and spin degrees of freedom.[4] Moreover, owing to their chemical stability, low-cost synthesis, zero field-hysteresis and large value of the isothermal entropy change near room temperature,[5] the possible application in energy-efficient and environment-friendly magnetic-refrigeration[6–9] has drawn attention. The general formula of a manganite perovskite is $AMnO_3$, where A is a trivalent atom. Replacing the A site atom with a divalent atom with smaller or larger atomic radius, known as hole doping,[1] results in a change of unit cell volume which is often referred to as a change in chemical pressure inside the material. An increment of the unit cell volume corresponds to a decrease of the chemical pressure.[10] The chemical substitution process causes the replacement of some of the $Mn^{3+}$ ions with $Mn^{4+}$ ions to maintain charge equilibrium. This creates the $Mn^{3+}$-$O^{2-}$-$Mn^{4+}$ double-exchange interaction in the compound, which is responsible for the low temperature ferromagnetic phase of the material and hence determines the Curie temperature ($T_C$) of the material. Thus, varying the amount of divalent cation in the A site, enables to tune the phase transition temperature of the material.

Phase inhomogeneity and quenched disorder in hole-doped manganites often leads to Griffiths phase (**GP**) formation. In manganites, quenched disorder can arise from several sources, such as, bending of the Mn-O-Mn bond angle, size variation of cations by chemical substitution, J-T distortions, etc.[11] In 1969, R. Griffiths observed a nonanalytic behaviour of the magnetization above $T_C$ in a randomly diluted Ising ferromagnet, caused by the formation of ferromagnetic clusters above $T_C$.[12] In a GP material, some lattice sites are either vacant or occupied with spin-zero atoms.[11] This intermediate phase between the ferromagnetic (**FM**) and paramagnetic (**PM**) phases, is referred as the GP.[13] In $La_{1-x}Pb_xMnO_{3\pm y}$ (**LPMO**), one possible reason of magnetic disorder is the competition between ferromagnetic $Mn^{3+}$-$O^{2-}$-$Mn^{4+}$ double-exchange interaction and antiferromagnetic $Mn^{3+}$-$O^{2-}$-$Mn^{3+}$ (or $Mn^{4+}$-$O^{2-}$-$Mn^{4+}$) super-exchange interaction. The probability for the existence of the GP decreases with increasing $Pb^{2+}$ substitution, as it decreases the amount of disorder in the material. Therefore, with increasing $Pb^{2+}$ substitution the signature of the GP-singularity will be suppressed up to a chemical substitution level where the GP disappears. Similar behaviour was observed for hole-doped $La_{1-x}Sr_xMnO_3$, where the GP disappeared for x > 0.16.[14] According to the theoretical model of Bray[15] and Bray-Moore[16], the ferromagnetic clusters in the GP grow in size with decreasing temperature in a way that it creates an effective magnetic long-range order in the material.[17] One approach to understand the magnetic ordering, is to investigate the critical scaling behaviour and to evaluate the critical exponents ($\beta$, $\gamma$ and $\delta$) as mentioned by Arrott-Naokes[18] and Kouvel-Fished.[19] Different theoretical models exist predicting different values for the critical exponents; mean-field ($\beta$ = 0.5, $\gamma$ = 1, $\delta$ = 3), three dimensional (3D)- Heisenberg ($\beta$ = 0.365, $\gamma$ = 1.336, $\delta$ = 4.8), 3D-Ising ($\beta$ =

0.325, $\gamma$ = 1.24, $\delta$ = 4.82) and tricritical mean-field ($\beta$ = 0.25, $\gamma$ = 1, $\delta$ = 5).[20] Previously, the magnetic ordering of $La_{0.7}Pb_{0.3}MnO_3$ has been explained in the framework of the long-range, mean-field like interactions at lower field and the observed critical exponents at higher field were reported to be more close to the 3D short-range interaction model.[21,22] Moreover, independent of magnetic field, a short-range, 3D-Heisenberg like critical behaviour was observed for $La_{0.9}Pb_{0.1}MnO_3$.[23] While performing the critical analysis it is important to perform the analysis in the critical region (~$T_C \pm 0.01T_C$) as described by Arrott and Noakes[18] as well as to compensate for the demagnetization effect and to use the intrinsic magnetic field in the analysis.[24] There are a large number of publications[25,26,35,36,27–34] on manganite materials reporting on critical scaling analysis without considering the two above mentioned criterions, which raises questions on the reliability of such analysis. Some other works[21–23,37–43] on Pb-substituted $LaMnO_3$ showed a PM-FM phase transition without any intermediate GP and $T_C$ was shown to increase with increasing $Pb^{2+}$ ionic substitution.

In this work, the $La^{3+}$ ion (1.216 Å) has been substituted by the larger $Pb^{2+}$ ion (1.35 Å) in the $LaMnO_3$ compound. The change of magnetic properties and evolution of the GP as a function of chemical substitution have been investigated. With critical scaling analysis (using intrinsic magnetic field in the analysis), the type of critical behaviour in the LPMO samples have been determined and compared with theoretical models. The magnetocaloric effect of the LPMO samples has also been studied. To the best of our knowledge, this is the first report of the evolution of the GP in the Pb-substituted $LaMnO_3$ system.

## 2. EXPERIMENTAL DETAILS

Polycrystalline LPMO samples were synthesized by using the solid-state reaction route. Stoichiometric amounts of $La_2O_3$, PbO and $MnCO_3$ powders were mixed together and calcinated at 1473 K for 24 h in Pt crucible with intermediate heating and grinding. All high-temperature treatments were performed at ambient atmosphere with a programmed heating and cooling rate of 50 K/hour. The samples were characterized using X-ray powder diffraction (XRPD) at 295 K by using Cu-$K_\beta$ radiation (Bruker D8 Advance diffractometer) with a step size of 0.013° (counting time was 15 s per step). The oxidation states of the samples were determined by using X-ray photoelectron spectroscopy (XPS). A "PHI Quantera II" system with an Al-K$\alpha$ X-ray source and a hemispherical electron energy analyzer with pass energy of 26.00 eV were used to collect the XPS spectra. The magnetic measurements were performed using a superconducting quantum interference device (SQUID) based Quantum Design magnetometer (MPMS) in the temperature range from 390 K to 5 K with a maximum field of 5 T.

# 3. RESULTS AND DISCUSSIONS

## 3.1. Structural and chemical properties

The single-phase formation of the LPMO samples has been confirmed by the Rietveld refinement of the XRPD patterns (*Figure 1 (a)*). Three structural models; orthorhombic (Pnma), rhombohedral (R-3c) and monoclinic (P2$_1$/n), were tested on the samples by using the Fullprof program[44] and the best fitting was observed for the rhombohedral structure in the R-3c space group. The extracted lattice parameters and average A-site and B-site radii, $\langle r_A \rangle$ and $\langle r_B \rangle$, calculated from Shannon ionic radii ($r_{La}^{3+}$ = 1.216 Å, $r_{Pb}^{3+}$ = 1.35 Å, $r_{Mn}^{3+}$ = 0.645 Å, $r_{Mn}^{4+}$ = 0.53 Å), are summarized in *Table 1*. With increasing $Pb^{2+}$ concentration, the lattice volume decreases which is explained by the decrease of the B-site ionic radius as $Mn^{3+}$ ions are replaced by smaller $Mn^{4+}$ ions. At the same time the increment of the Mn-O-Mn bond angle and the decrement of Mn-O bond length were observed. The shifting (*Figure 1 (b)*) of the highest intensity peak (110) towards larger angle and the merging of the peaks (*Figure 1 (c) and (d)*) indicate lattice contraction with increasing chemical substitution.

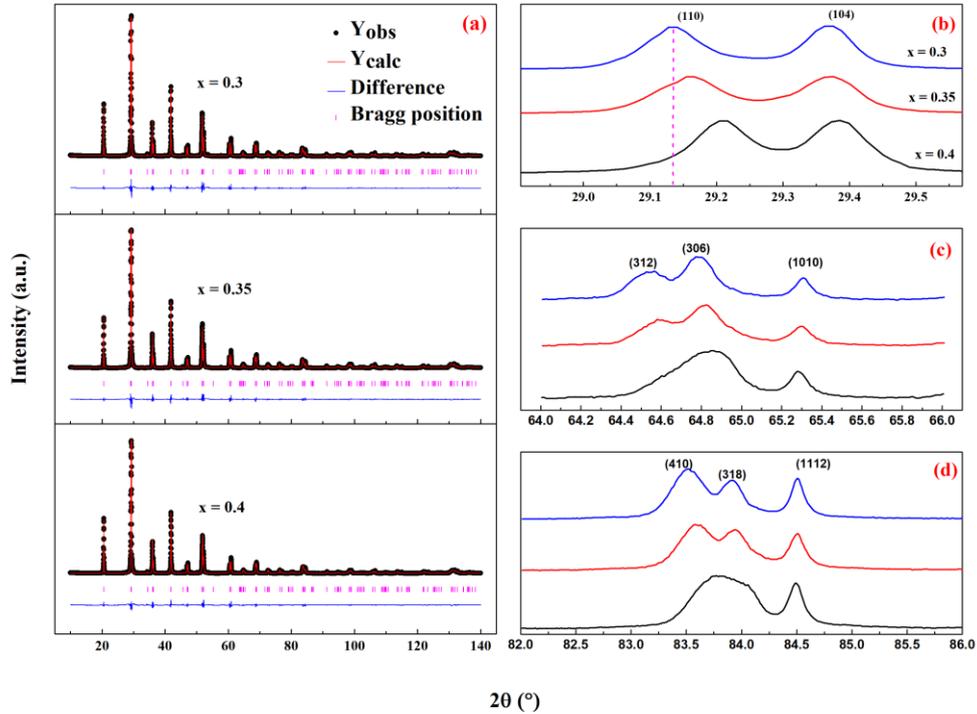

*Figure 1 (a) Rietveld refinement of XRPD patterns of LPMO samples. (b) right- Shifting of the (110) peak with increasing $Pb^{2+}$ substitution. (c) and (d) peak merging with increasing $Pb^{2+}$ substitution.*

*Table 1 Structural parameters for LPMO from Rietveld refinement of XRPD spectra.*

| $x$ | | 0.3 | 0.35 | 0.4 |
|---|---|---|---|---|
| $\langle r_A \rangle$ (Å) | | 1.256 | 1.263 | 1.270 |
| $\langle r_B \rangle$ (Å) | | 0.611 | 0.605 | 0.599 |
| $a$ (Å) | | 5.5338(1) | 5.5286(1) | 5.5213(1) |
| $c$ (Å) | | 13.3973(3) | 13.4006(3) | 13.4046(3) |
| $V$ (Å$^3$) | | 355.3(1) | 354.7(1) | 353.9(1) |
| Mn-O (Å) | | 1.968(4) | 1.966(4) | 1.962(4) |
| Mn-O-Mn (°) | | 164.16(3) | 164.32(3) | 165.61(3) |
| **Main fitting parameters** | $R_p$ | 4.38 | 4.61 | 4.71 |
| | $R_{wp}$ | 6.21 | 6.31 | 6.38 |
| | $R_B$ | 4.32 | 4.12 | 5.02 |

The ionic state of Mn plays an important role in the lattice distortion. To check the manganese ionic state, the Mn-2p peak has been analysed by XPS (*Figure 2*). The observed amount of Mn$^{4+}$ for the LPMO samples with increasing Pb$^{2+}$ substitution was 30.06 (±1.54) %, 34.05 (±1.66) % and 39.38 (±1.61) %, which can be compared to their expected amounts of 30%, 35% and 40%, respectively. The spin-orbit splitting between Mn2p$_{3/2}$ and Mn2p$_{1/2}$ was ~ 11.7 eV for all samples. The absence of any satellite peak confirms that there are no Mn$^{2+}$ ions in the LPMO compounds.[45]

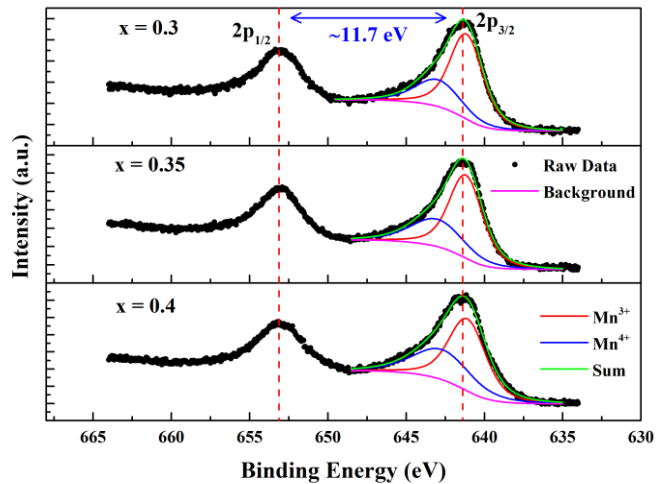

*Figure 2 Mn2p XPS spectra for the different LPMO samples.*

## 3.2. Magnetic properties

*Figure 3* shows the temperature dependence of the magnetization for the LPMO compounds. A PM-FM phase transition was observed for all three samples.

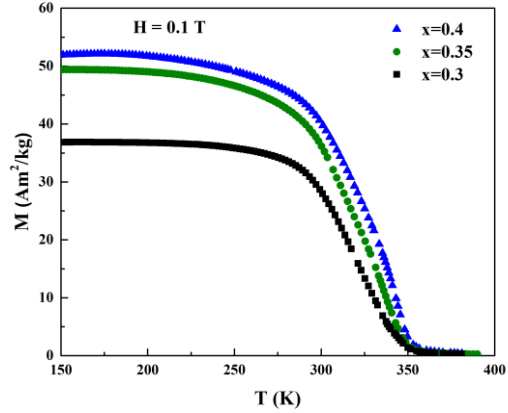

*Figure 3* *Temperature dependence of the magnetization for LPMO samples with different levels of $Pb^{2+}$ substitution.*

In the double-exchange (DE) model,[46,47] the magnetic interaction is attributed to the orbital overlap between of 3d- and 2p-orbitals of Mn- and O-atoms, respectively. The charge carrier bandwidth $W$ is a measure of the orbital overlap,[48] which is defined as,

$$W \propto \frac{\cos(1/2(\pi - \langle Mn-O-Mn \rangle))}{d_{Mn-O}^{3.5}}, (1)$$

where $\langle Mn - O - Mn \rangle$ and $d_{Mn-O}$ are the bond angle and bond length, respectively. From *Equation* (1), the highest value of $W$ is observed for the $x = 0.3$ sample and it decreases by ~6% and ~7% for the $x = 0.35$ and 0.4 samples, respectively. For some perovskite manganites, it has been found that the increase/decrease of $T_C$ has a direct proportionality with the increase/decrease of $W$ [48]. However, for LPMO, $T_C$ increases with decreasing $W$ (cf. Figure 3 and $T_C$ values shown in Table 2 obtained from critical scaling analysis). The effect of correlated disorder on the magnetism in double exchange systems has been theoretically investigated [49]. In particular it was discussed how disorder introduced by chemical substitution affects the hole density and $T_C$. From stoichiometry, the hole density will be similar to the concentration of the dopant. The theoretical results show that $T_C$ increases with increasing hole density up to a certain level after which $T_C$ decreases on further increase of the hole density [49]. Thus, for the LPMO compounds studied in this work the hole density ($Mn^{4+}$ concentration) is below the level where $T_C$ reaches its maximum value.

To understand the magnetic behaviour in more detail, the variation of the inverse DC-susceptibility with temperature was studied (*Figure 4(a)*). For a pure PM-FM transition, in the paramagnetic region, the magnetic susceptibility ($\chi$) follows the Curie-Weiss law,

$$\chi = \frac{C}{T - \theta_P}, (2)$$

Where, $C$ is the Curie constant, $T$ is temperature and $\theta_P$ is the Curie-Weiss temperature. For LPMO, deviations from the expected linear dependence between $1/\chi$ and $T$ are observed. A deviation from a linear dependence can also be expected for ferrimagnets. But for ferrimagnets, the tangent in the paramagnetic region at temperatures $T \gg \theta_P$ of the $1/\chi$ vs. $T$ curve should have negative intercept with the temperature axis,[50] which is not observed for LPMO.

Based on the original work of Griffiths[12] and its following works,[14,51–54] the magnetic behaviour of LPMO resembles that of a GP material. According to this model the magnetic system at temperatures $T > T_C$ can be described as a disordered system with randomly distributed FM clusters embedded in a PM background.[54] This GP can be characterized by

$$\chi^{-1} \propto (T - T_C^R)^{1-\lambda}, (3)$$

where $\lambda$ is a constant and $T_C^R$ is the transition temperature associated with the GP.[53,55] There is also another characteristic temperature ($T_G$), above which the pure FM behaviour (following the Curie-Weiss law) is observed.[14] The GP behaviour is observed in the temperature regime $T_C^R \leq T \leq T_G$.[55] The value of $\lambda$ is a measure of the deviation from a "pure" ferromagnetic behaviour. Therefore, a higher value of $\lambda$ corresponds to a more disordered state. The A-site chemical substitution and field dependence of $\lambda$ is shown in *Figure 4(b)*. The most disordered state is observed for the $x = 0.3$ sample and with increasing A-site chemical substitution the effect of disorder is suppressed. Also, with increasing magnetic field, the GP singularity is suppressed.

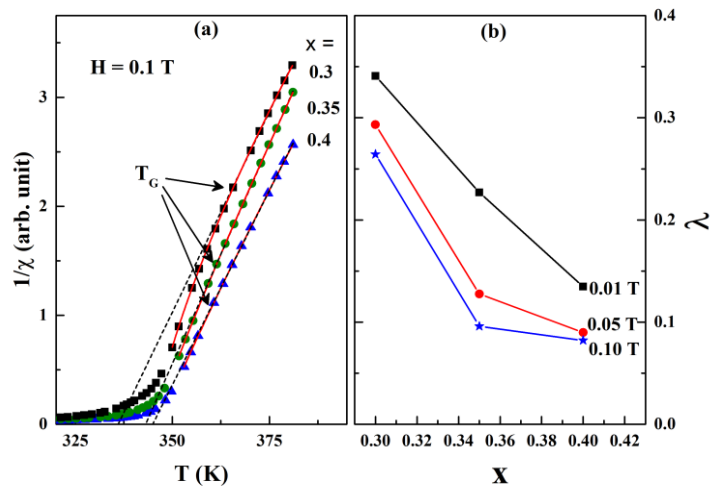

**Figure 4 (a)** Temperature dependence of the inverse susceptibility. The red solid lines correspond to fits to *Equation* (3). The black dotted lines represent Curie-Weiss law fits. **(b)** Variation of λ with $Pb^{2+}$-substitution for different magnetic fields. The solid lines are guide to the eye.

### 3.3. Scaling analysis

The critical scaling analysis has been performed using the intrinsic magnetic-field ($H_i$) to avoid the effect of the demagnetization field.[24] The magnetic transition was analysed using Arrott-plots[56] and the Banerjee-criterion[57] as shown in *Figure 5*. Absence of any negative slope in the plot confirms the second order magnetic phase transition for all three materials (results for $x = 0.35$ and $0.3$ are not shown here, they are similar to the $x = 0.4$ results).

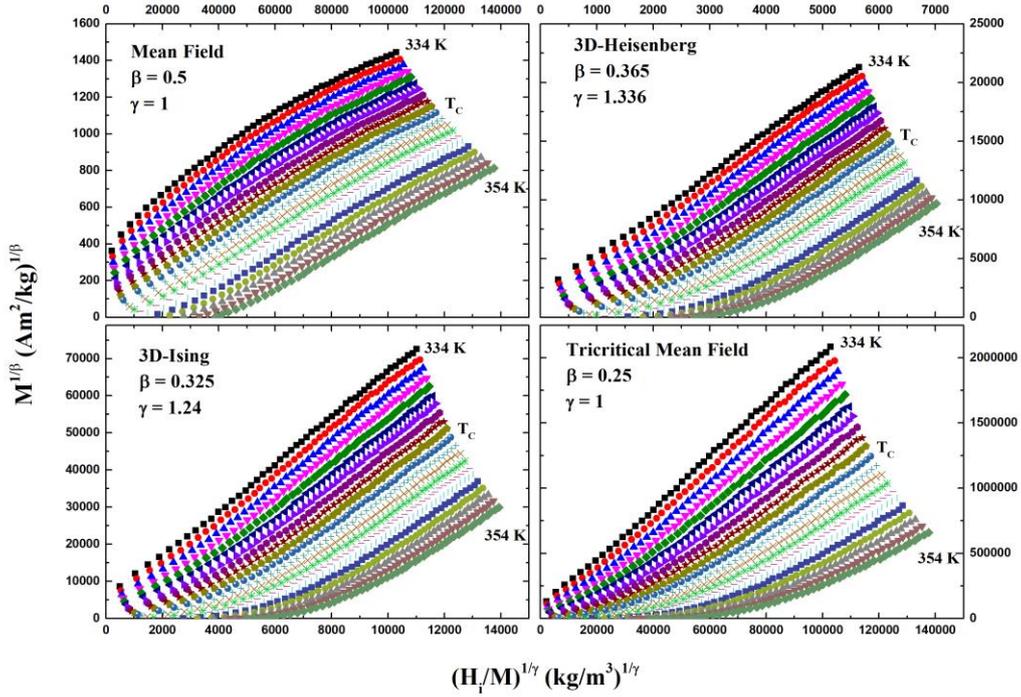

*Figure 5 Modified Arrott plots with mean-field, 3D-Heisenberg, 3D-Ising and tricritical mean-field models for x = 0.4 sample.*

The critical behaviour of a material with a second order phase transition is defined with a set of critical exponents near the $T_C$. The critical exponents $\beta$, $\gamma$ and $\delta$ correspond to the spontaneous magnetization ($M_{sp}$), initial susceptibility ($\chi$) and the magnetization isotherm at $T_C$, respectively, and are described with the following set of equations,[58]

$$M_{sp}(0,T) \propto (-\varepsilon)^{\beta}; \; T < T_C \, , (4)$$

$$\frac{1}{\chi}(0,T) \propto \varepsilon^{\gamma}; \; T > T_C \, , (5)$$

$$M(H,T) \propto H^{1/\delta}; \; T = T_C \, , (6)$$

Where, $\varepsilon = \frac{T-T_C}{T_C}$. The exponents can according to Widom scaling be related as,[20,59]

$$\delta = 1 + \frac{\gamma}{\beta}, (7)$$

The value of $T_C$ is extracted from the Arrott plots as described by D. Kim et al.[24] and are listed in *Table 2*. The FM-PM transition is expected at $T_C$, but the GP does not allow the material to leave its FM-state until $T_G$. As the GP-region is near the critical region, it limits the use of *Equations* (4) and (5). To avoid any effect of the GP in the critical analysis, instead of using *Equations* (4) and (5) the static scaling hypothesis has been used,[24] according to which,

$$M(\varepsilon, H)\varepsilon^{-\beta} = f_\pm(H\varepsilon^{-(\beta+\gamma)}), (8)$$

where, $f_+$ is for $\varepsilon > 0$ and $f_-$ is for $\varepsilon < 0$. Using *Equation* (7), the equation of state can be expressed as,

$$M(\varepsilon, H)\varepsilon^{-\beta} = f_\pm(H\varepsilon^{-(\beta\delta)}), (9)$$

Using *Equation* (6), the value of $\delta$ has been extracted for the different samples and then using *Equation* (9) the value of $\beta$ has been found from the best data-collapse, as is shown in *Figure 6(a)*. The values of the extracted critical exponents are listed in *Table 2*.

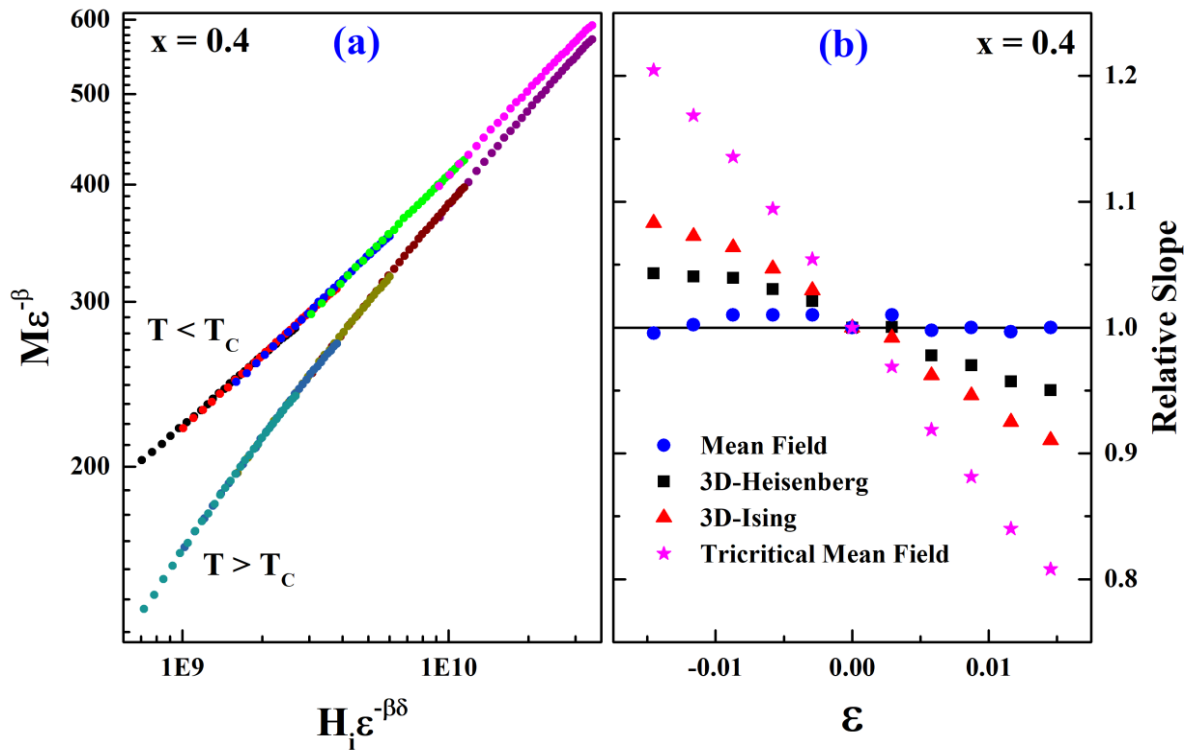

***Figure 6** (a) Scaling plot for x=0.4 sample using the critical exponent values from **Table 2**, where the absolute value of ε has been used. (**b**) Relative slope of Arrot plot isotherms for different models. $T_C$ ± 5 K has been used for this comparison.*

*Table 2 Values of critical exponents.*

| Compounds | | β | δ | γ (calculated) | $T_C$ (K) (from scaling) | Reference |
|---|---|---|---|---|---|---|
| x | 0.3 | 0.51(1) | 3.04(3) | 1.04(4) | 326(1) | This work |
|   | 0.35 | 0.51(1) | 2.91(2) | 0.97(3) | 332(1) | This work |
|   | 0.4 | 0.50(1) | 3.19(1) | 1.1(1) | 344(1) | This work |
| Mean-Field (MF) | | 0.5 | 3.0 | 1 | - | [20] |
| 3D-Heisenberg | | 0.365 | 4.8 | 1.336 | - | [20] |
| 3D-Ising | | 0.325 | 4.82 | 1.24 | - | [20] |
| Tricritical MF | | 0.25 | 5.0 | 1 | - | [20] |

The extracted values of $\beta$, $\delta$ and $\gamma$, resemble the values valid for the mean-field model. In order to further establish which of the four models (mean-field, 3D-Heisenberg, 3D-Ising and tricritical mean-field) best fits the results of the LPMO samples, modified Arrott plots are shown in *Figure 5* for the $x$ =0.4 sample (the behaviour of the other samples is similar). For the best fitted model, the critical isotherm at $T = T_C$ should pass through the origin and all other isotherms near $T_C$ should be parallel to the critical isotherm.[24] To confirm this parallel-isotherm feature near the $T_C$, the slopes for the $x$ =0.4 sample with respect to the critical isotherms are presented in *Figure 6(b)* for the different models. From this relative slope plot (the behaviour of the three samples is similar), the observed best model is the mean-field model for the LPMO samples.

### 3.4. Magnetocaloric Effect (MCE) Analysis

To investigate the MCE indirectly,[6] the magnetic entropy change ($\Delta S_M$) has been calculated from isothermal magnetization curves. At constant temperature $T$, changing the magnetic field from 0 to $\mu_0 H_{max}$, $\Delta S_M$ is defined as,[5,60]

$$\Delta S_M(T, \mu_0 H) = S_M(T, \mu_0 H) - S_M(T, 0) = \int_0^{\mu_0 H_{max}} \left(\frac{\partial S}{\partial(\mu_0 H)}\right)_T d(\mu_0 H) , (10)$$

Using Maxwell's thermodynamic equation,

$$\left(\frac{\partial S}{\partial(\mu_0 H)}\right)_T = \left(\frac{\partial M}{\partial T}\right)_{\mu_0 H} , (11)$$

$\Delta S_M$ can be redefined as,

$$\Delta S_M(T, \mu_0 H) = \int_0^{\mu_0 H_{max}} \left(\frac{\partial M}{\partial T}\right)_{\mu_0 H} d(\mu_0 H), \quad (12)$$

Using this relation, the values of $\Delta S_M(T)$ were calculated by measuring the magnetic moment with increasing magnetic field up to 5 T for a temperature range of 280 K to 370 K. In *Figure 7* the magnetic entropy changes for different $\mu_0 H_{max}$ shows a maximum value at a temperature around the PM-FM phase transition region.

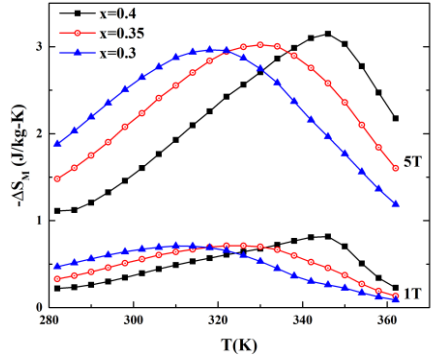

*Figure 7* Isothermal entropy change $\Delta S_M(T)$ for $\mu_0 H_{max}$ = 1 T and 5 T.

Another important parameter for the application of the MCE is the relative cooling power (RCP), which decides the temperature span in which the magnetic refrigerator can work. The RCP is defined as,[9]

$$RCP = -\Delta S_M^{max} \times \Delta T_{FWHM}, \quad (13)$$

where $-\Delta S_M^{max}$ is the maximum value of the isothermal entropy change and $\Delta T_{FWHM}$ is the full width at half maximum of the isothermal entropy change with respect to temperature. The calculated RCP values (258 J/kg, 244 J/kg and 223 J/kg for $x$=0.3, 0.35 and 0.4, respectively) decrease with increasing $Pb^{2+}$ substitution, which makes the $x$=0.3 sample best suited for MCE applications.

## 4. SUMMARY and CONCLUSIONS

The magnetic properties of LPMO, including a GP-singularity and critical scaling behaviour are reported. A PM-FM phase transition was observed for all samples. The phase transition temperature ($T_C$) increases with increasing A-site chemical substitution. However, opposed to predictions of the double-exchange model, $T_C$ decreases with increasing charge carrier bandwidth. All LPMO samples showed a GP-singularity at low magnetic field and this singularity was suppressed by increasing external magnetic field and increasing A-site chemical substitution. From the Arrott-plots[56] and the Banerjee-criterion[57] a second order magnetic phase transition was confirmed for all three compounds.

Using the Arrott-Naokes[18], Kouvel-Fished[19] and Widom[20,59] rules the values of universal critical exponents ($\beta$, $\gamma$ and $\delta$) have been calculated and they are very close to the mean-field model values. The observed mean-field like long-range interaction is attributed to the formation of ferromagnetic clusters[21] in the presence of the GP. Long-range ferromagnetic order is established as the result of the interaction between ferromagnetic clusters. Using a renormalization group approach, Fisher et al.,[61] derived the critical exponents for a system where the decay of the long-range interaction was described as $J(r) \propto \frac{1}{r^{d+\sigma}}$, where $d$ is the dimension of the system and $\sigma$ is an exponent describing the decay of the interaction potential. The transition from mean-field to short-range critical behaviour was discussed in terms of the variable $\vartheta = 2\sigma - d$ and it was found that mean-field critical exponents were obtained for $\vartheta < 0$. From our critical scaling analysis, it can therefore be concluded that the interaction potential describing the interaction between ferromagnetic clusters in the LPMO samples is described by an exponent $\sigma < 1.5$. The magnetocaloric analysis showed a small change in the maximum value of $\Delta S_M$ (isothermal entropy change) with A-site chemical substitution.

## ACKNOWLEDGEMENTS

The Swedish Foundation for Strategic Research (SSF, contract EM-16-0039) supporting research on materials for energy applications is gratefully acknowledged. Infrastructural grants by VR-RFI (#2017-00646_9) and SSF (contract RIF14-0053) supporting accelerator operation are gratefully acknowledged. Financial support by FITC HF RAS through project No. 45.22 (grant AAAA18-118012390045-2) is gratefully acknowledged. The authors are thankful to Daniel Hedlund for fruitful discussions and proofreading. The authors are thankful to Sanchari Chakraborti for help in Arrott-plot data plotting.